\documentclass[prd,12pt]{revtex4-1}

\usepackage{amsmath}
\usepackage{commath}
\usepackage{graphicx}
\usepackage{amssymb}
\usepackage{subfigure}
\usepackage{bm}
\usepackage{slashed}
\usepackage{tikz}
\usepackage{braket}
\usepackage{multirow}
\usepackage[labelsep=space]{caption}
\usepackage[normalem]{ulem}

\def\process{\mbox{$B \to K^{*} \mu^+ \mu^-$ }}

\begin{document}
\title{Effect of $c\overline{c}$ resonances in the branching ratio and forward-backward asymmetry of the decay \process}

\author{Mohammad Ahmady}
\email{mahmady@mta.ca}
\affiliation{\small Department of Physics, Mount Allison University, \mbox{Sackville, New Brunswick, Canada, E4L 1E6}}
\author{Dan Hatfield}
\email{dhatfield@mta.ca}
\affiliation{\small Department of Physics, Mount Allison University, \mbox{Sackville, New Brunswick, Canada, E4L 1E6}}
\author{S\'{e}bastien Lord}
\email{esl8420@umoncton.ca}
\affiliation{D\'epartement de Math\'ematiques et Statistique, Universit\'{e} de Moncton, \mbox{Moncton, New Brunswick, Canada, E1A 3E9}}
\author{Ruben Sandapen}
\email{rsandapen@mta.ca}
\affiliation{\small Department of Physics, Acadia University, Wolfville, Nova-Scotia, Canada, B4p 2R6}
\affiliation{\small Department of Physics, Mount Allison University, \mbox{Sackville, New Brunswick, Canada, E4L 1E6}}

\begin{abstract}
We compute the branching ratio and forward-backward asymmetry ($A_{FB}$) distribution for the rare dileptonic decay $B \to K^* \mu^+ \mu^-$ for the full range of $q^2$, the dimuon mass squared, region. For the required form factors, we use nonperturbative inputs as predicted by the anti-de Sitter (AdS)/QCD correspondence. When using the Breit-Wigner model with momentum-dependent decay constants to account for the $\psi$ and $\psi^\prime$ resonance effects in the nonresonance region of the spectrum,  we find our predictions to be in better agreement with the experimental data for the branching ratio.
\end{abstract}


\maketitle

\section{Introduction}

The rare decay $B \to K^* \mu^+ \mu^-$ has recently been attracting much attention  from both the experimental \cite{Aaij:2014pli,Aaltonen:2011ja,Lees:2012tva,Wei:2009zv,Aaij:2013qta,Aaij:2012cq,Aaij:2013iag,LHCb:2015dla,Khachatryan:2015isa} and theoretical \cite{Altmannshofer:2013foa,Descotes-Genon:2013wba,Hurth:2013ssa,Descotes-Genon:2013vna,Gauld:2013qba,Buras:2013qja,Gauld:2013qja,Hambrock:2013zya,Khodjamirian:2010vf,Bharucha:2010im,Datta:2013kja,Alok:2009tz} sides due to the various observables associated with this decay that are susceptible to reveal new physics (NP). In particular, reference \cite{Descotes-Genon:2013wba} has brought to light an overall tension between the Standard Model predictions and the experimental data and has suggested that a modification to the $C_{7,9}$ Wilson coefficients could resolve this tension.

To investigate signals of NP, one usually focuses on the region of the spectrum away from $\psi$ and $\psi^\prime$ resonances where short-distance (SD) interactions, as represented by Figs.\ \ref{fig:feynmanndiagrams_SD_penguin} and \ref{fig:feynmandiagrams_SD_box}, are dominant.  Experimentally, the $q^2$ region around the above two resonances are subtracted from the dileptonic spectrum. However, a careful analysis of \process observables should consider the long-distance effects of the resonances in the SD dominated region.  In this paper, we take into account the narrow resonance effects in the nonresonance region when calculating the differential decay rate and the forward-backward asymmetry in this decay.  In doing so, we use a Breit-Wigner model for the resonances with momentum-dependent decay constants\cite{Ahmady:1995st}.  We note that the latter model fits the data on photoproduction and leptonic width of $\psi$ and $\psi^\prime$ simultaneously\cite{Terasaki:1981ts}, is used for exclusive \process for the first time.  The effects of broad resonances, using quark hadron duality, are considered in Ref. \cite{Khodjamirian:2010vf,Lyon:2013gba}.

In a previous paper \cite{Ahmady:2014sva}, we have computed the full set of 7 independent $B \to K^*$ transition form factors. At low-to-intermediate $q^2$, we used light-cone Sum Rules with AdS/QCD distribution amplitudes (DAs)\cite{Ahmady:2013cva}. These DAs are derived from the holographic AdS/QCD light-front wavefunction for $K^*$\cite{deTeramond:2008ht,Brodsky:2014yha}. We have fitted these with form factor predictions at high $q^2$ from lattice QCD . In this work, the same method for the derivation of the form factors with updated inputs (B meson decay constant $f_B$ and b-quark mass $m_b$) are used to calculate the differential branching ratio and forward-backward asymmetry in the decay \process.

We find that including the resonance effects improves the agreement of our predictions with the LHCb data \cite{Aaij:2014pli} and the latest CMS data \cite{Khachatryan:2015isa} on the differential branching ratio.  As for $A_{FB}$, it seems that the inclusion of the resonances hardly changes our prediction for dimuon mass squared below the first resonance. Finally, we find that a negative shift in the Wilson coefficient $C_9$ enhances the agreement with the data for the differential branching ratio and the $A_{FB}$  at $q^2$ below the first $c\bar c$ resonance.

%%%%%%%%%%%%%%%%%%%%%%%%%%%%%%%%%%%%%%%%%%%%%%%%%%%%%%%%
\section{Differential branching ratio with resonances} %
%%%%%%%%%%%%%%%%%%%%%%%%%%%%%%%%%%%%%%%%%%%%%%%%%%%%%%%%

\begin{figure}
\centering
\subfigure[\mbox{ }SD penguin diagram]{\includegraphics[width=0.3\textwidth]{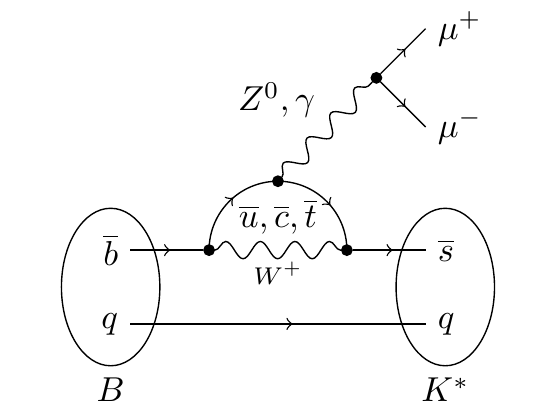}
\label{fig:feynmanndiagrams_SD_penguin}}
\subfigure[\mbox{ }SD box diagram]{\includegraphics[width=0.3\textwidth]{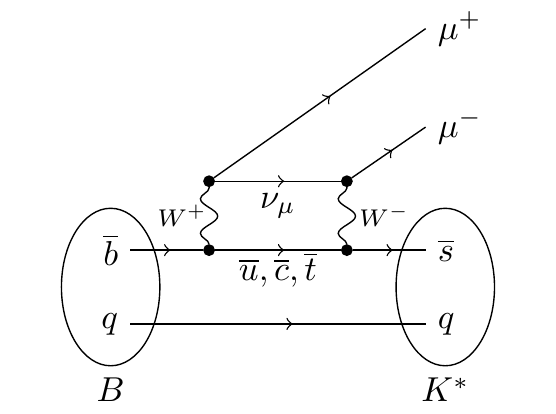}
\label{fig:feynmandiagrams_SD_box}}
\subfigure[\mbox{ }LD resonance diagram]{\includegraphics[width=0.3\textwidth]{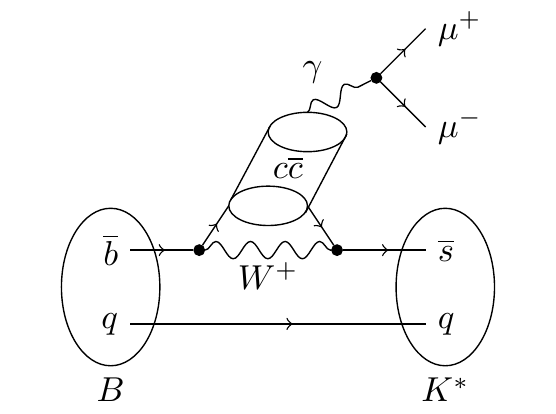}
\label{fig:feynmandiagrams_LD_ccbar}}
\caption{Feynman diagrams of the principal contributions to the \process decay.}
\label{fig:feynmandiagrams}
\end{figure}

In our previous paper \cite{Ahmady:2014sva}, we calculated the differential branching ratio for \process without considering the effects of resonances. The inclusion of $\psi$ and $\psi^\prime$ resonances, as illustrated in Fig.\ \ref{fig:feynmandiagrams_LD_ccbar},  is obtained by modifying $C_9^{\rm eff}$ with an additional term $C_9^{\rm res}$ which, using the Breit-Wigner model, can be written as \cite{PhysRevD.39.1461, Lim1989343}:
\begin{equation}
C_9^{\rm res} = (3C_1(\mu) + C_2(\mu)) \frac{16\pi^2}{9} \left(\frac{f^2_\psi/m_\psi^2}{m_\psi^2 - q^2 - im_\psi\Gamma_\psi} + (\psi \to \psi')\right)\; ,
\end{equation}
where $C_1$ and $C_2$ are the Wilson coefficients corresponding to the current-current operators $O_1$ and $O_2$ evaluated at scale $\mu\sim m_b$ and $f_{\psi^{(')}}$ and $\Gamma_{\psi^{(')}}$ are the decay constant and total width of the $c\bar c$ resonance $\psi^{(')}$, respectively.  We use the same convention for the effective operators as in reference \cite{Beneke:2001at} and the following definition for vector meson decay constant:
\begin{equation}
\braket{0|\bar c\gamma_\mu c|V}=f_V\epsilon_\mu \; .
\end{equation}
Since $\psi$ and $\psi^\prime$ resonances are off mass-shell for $q^2$ different from $m_{\psi^{(')}}^2$ in \process, we need to consider the  $q^2$-dependence of their decay constants \cite{Ahmady:1995st}:
 \begin{equation}
 f_V(q^2) = f_V(0)\left(1+\frac{q^2}{c_V}[d_V - h(q^2)]\right) \hspace{1cm} (V = \psi, \psi')
 \label{eq:runningfv}
 \end{equation}
 with the $h$ function being related to the imaginary part of the quark-loop diagram:
 \begin{equation}
 h(q^2) = \frac{1}{16\pi^2r}\left(-4 - \frac{20r}{3}+4(1+2r)\sqrt{1-\frac{1}{2}}\arctan\frac{1}{\sqrt{1-\frac{1}{r}}}\right)
 \label{eq:quarkloop}
 \end{equation}
where $r = q^2/4m_q^2$ for $0< q^2 < 4m_q^2$. $m_q$ is the effective quark mass and assuming that the vector mesons are weakly bound systems of a quark and an antiquark, we take $m_q= m_V/2$. As a result, Eq.\ \ref{eq:quarkloop}, defined for $0< q^2 <4m_q^2$, is an interpolation of $f_V$ from the experimental data on $f_V(0)$ (from photoproduction) and $f_V (m_V^2)$ (from leptonic width) based on a quark-loop diagram. We assume $f_V(q^2) = f_V(m_V^2)$ for $q^2 > m_V^2$. The numerical values of the parameters $c_V$ and $d_V$ in Eq.\ \ref{eq:runningfv} are given in Tab.\ \ref{table_runningfv} \cite{Ahmady:1995st}.

 \begin{table}
 	\begin{tabular}{| c || c | c | c | c |}
 		\hline
 		$V$ & $f_V(0)$ & $f_V(m_V^2)$ & $c_V$ & $d_V$ \\
 		\hline\hline
 		$\psi$ & $0.54$ & $1.25$ & $0.54$ & $0.77$ \\
 		\hline
 		$\psi'$ & $0.043$ & $1.04$ & $0.043$ & $0.043$ \\
 		\hline
 	\end{tabular}
 	\caption{Parameters (in GeV-based units) used in the $q^2$ evolution of $f_V$.}
	\label{table_runningfv}
 \end{table}
 \begin{figure}
 	\includegraphics{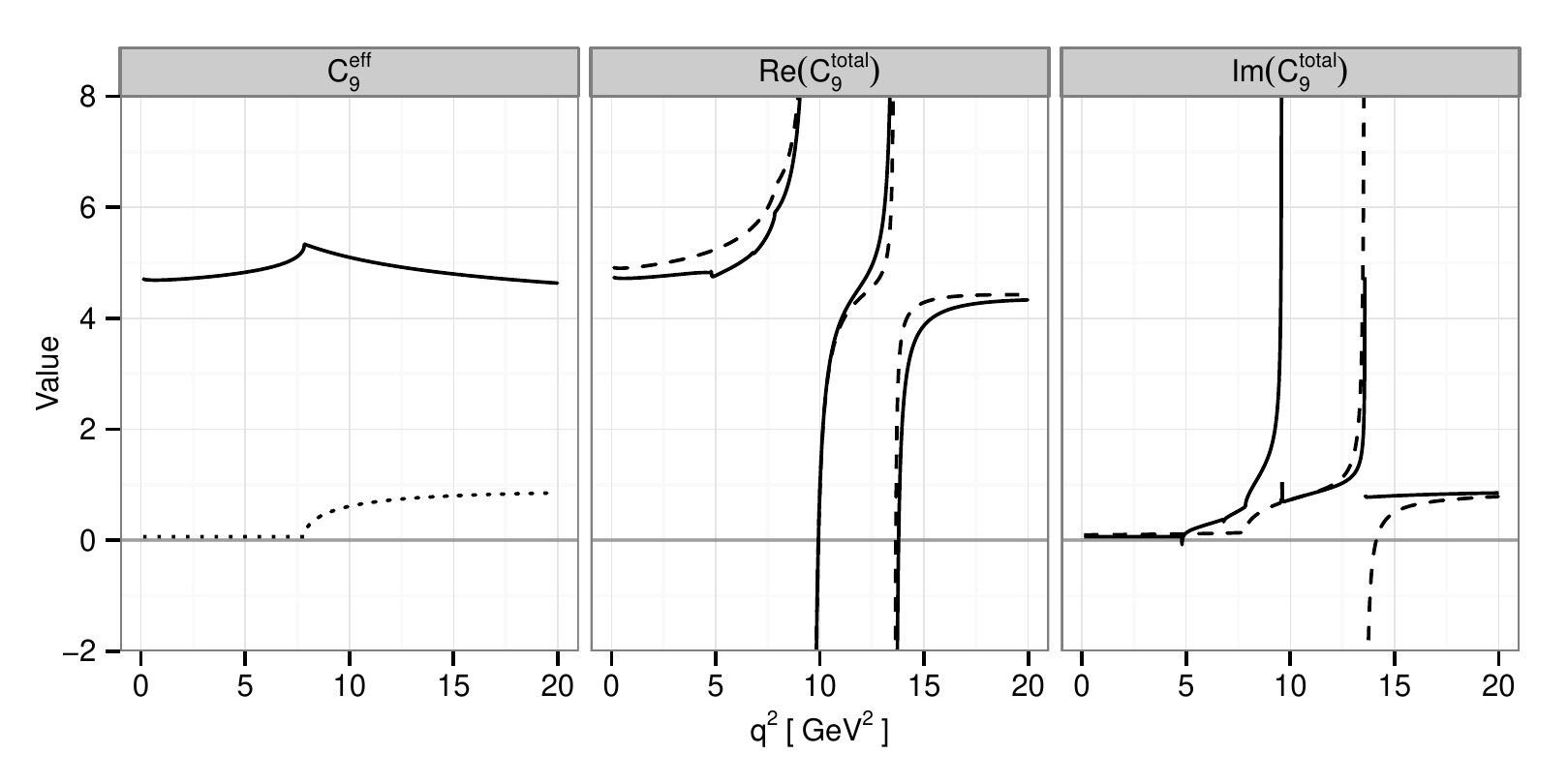}
 	\caption{Plots of $C_9^{\rm eff}$, $\Re(C_9^{\rm tot})$ and $\Im(C_9^{\rm tot})$ versus $q^2$. In the left plot, the solid curve is $\Re(C_9^{\rm eff})$ while the dotted curve is $\Im(C_9^{\rm eff})$. In the middle and right figures, the solid and dashed curves correspond to utilizing momentum-dependent and momentum-independent decay constants, respectively.}
 	\label{fig:C9}
 \end{figure}

The resonance contributions $C_9^{\rm res}$ augments the short-distance contributions $C_9^{\rm eff}$ in the effective Hamiltonian:
\begin{equation}
C_9^{\mbox{\tiny tot}} = C_9^{\mbox{\tiny eff}} -C_9^{\rm res} \; .
\label{Ctotal}
\end{equation}
The minus sign in Eq. \ref{Ctotal} is due to our choice of convention for the Wilson coefficients. The real and imaginary components of $C_9^{\rm tot}$ as a function of $q^2$ are shown in Fig.\ \ref{fig:C9}. To calculate the differential branching ratio including the resonance contributions, one should replace $C_9^{\rm eff}$ by $C_9^{\rm tot}$ in the differential branching ratio expression given in Ref. \cite{Ahmady:2014sva}.  

\begin{figure}
\includegraphics[width = 0.8\textwidth, height = 0.5\textwidth]{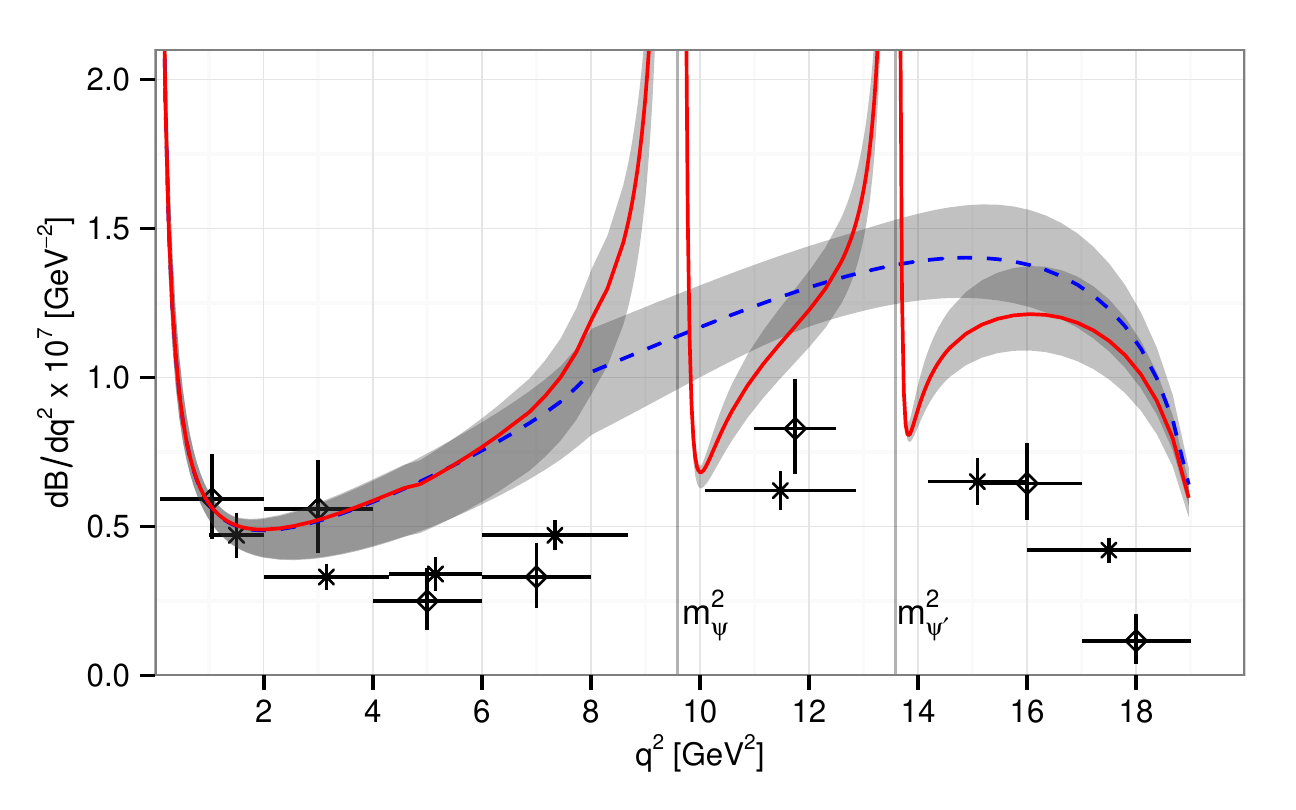}
\caption{The AdS/QCD prediction for the differential branching ratio of the \process decay, with (solid red) and without (dashed blue) resonances, as compared with the latest LHCb ($B^+ \to K^{*+} \mu^+ \mu^-$ (diamonds)) and CMS ($B^0 \to K^{*0} \mu^+ \mu^-$ (crosses)) data.  Note that this plot is qualitative and our predictions for each experimental bin for this observable are shown in Tab.\ \ref{tab:bin_BR} in Appendix A.}
\label{fig:BR_main}
\end{figure}

As for the 7 form factors which parametrize the $B\to K^*$ transition, they are calculated using AdS/QCD DAs \cite{Ahmady:2014sva} in conjunction with light-cone sum rules at low to intermediate $q^2$. For high $q^2$ values, we use the latest lattice data for $B\to K^*$ transition form factors \cite{Horgan:2013hoa}.  Note that we use the lattice results reported under ensemble f0062 as they correspond to finer lattice spacing. We use the following two-parameter form to fit the form factors obtained from AdS/QCD at low-to-intermediate $q^2$ and the lattice data at high $q^2$:
\begin{equation}
F(q^2) = \frac{F(0)}{1 - a\frac{q^2}{m_B^2} + b\frac{q^4}{m_B^4}}
\label{FFfit}
\end{equation}
The updated values for $F(0)$, $a$ and $b$ are given in Tab.\ \ref{tab:FFfits_val}.
\begin{table}
	\begin{tabular}{| c || c | c | c | c | c | c | c |}
		\hline
		$F$ & $A_0$ & $A_1$ & $A_2$ & $T_1$ & $T_2$ & $T_3$ & $V$\\
		\hline \hline
		$F(0)$ & $0.243$ & $0.244$ & $0.244$ & $0.258$ & $0.239$ & $0.157$ & $0.297$ \\ \hline
		$a$ & $1.618$ & $0.586$ & $1.910$ & $1.910$ & $0.525$ & $1.147$ & $1.934$ \\ \hline
		$b$ & $0.561$ & $-0.356$ & $1.498$ & $1.082$ & $-0.459$ & $-0.114$ & $1.089$ \\ \hline
	\end{tabular}
	\caption{Updated fit parameters for the seven independent $B \to K^{*}$ form factors used in Eq.\ \ref{FFfit}.}
	\label{tab:FFfits_val}
\end{table}
Our prediction for the differential branching ratio including the effects of the resonances $\psi$ and $\psi^\prime$ as obtained by using the above form factors is shown in Fig.\ \ref{fig:BR_main} where we compare with the latest data from LHCb \cite{Aaij:2014pli} and CMS \cite{Khachatryan:2015isa}. Our numerical results are calculated with the input parameters given in Tab.\ \ref{tab:inputs} and the Wilson coefficients tabulated in Tab.\ \ref{tab:Wilson_coef}. Fig.\ \ref{fig:BR_main} clearly shows the effect of including resonances with the momentum-dependent decay constant on our prediction of the differential branching ratio.

%%%%%%%%%%%%%%%%%%%%%%%%%%%%%%%%%%%%%
\section{Forward backward asymmetry}%
%%%%%%%%%%%%%%%%%%%%%%%%%%%%%%%%%%%%%
The forward-backward asymmetry distribution in dileptonic rare \process decay is defined as:
\begin{equation}
\frac{dA_{\rm FB}}{dq^2} \equiv 
\frac{1}{d\Gamma/dq^2} \left(\,
\int_0^1 d(\cos\theta_\ell) \,\frac{d^2\Gamma}{dq^2 
	d\cos\theta_\ell} - \int_{-1}^0 d(\cos\theta_\ell) \,\frac{d^2\Gamma}{dq^2 
	d\cos\theta_\ell} \right) \;
\end{equation}
where $\theta_\ell$ is the angle between the positive muon and the line of flight of $K^*$ in the $\mu^+ \mu^-$ rest frame.  This distribution to next-to-leading order(NLO) accuracy in $\alpha_s$ is given by \cite{Beneke:2001at}:
\begin{align}
\label{AFB}
\frac{\dif A_{FB}}{\dif q^2} = & -\frac{1}{\dif \Gamma / \dif q^2} \frac{G_F^2 \envert{V_{ts}^* V_{tb}}^2}{128 \pi^3} m_B^3 \lambda(q^2, m_{K^*}^2)^2 \left( \frac{\alpha_{em}}{4 \pi} \right)^2 C_{10} A_1(q^2) V(q^2) \nonumber \\
& \times \Re \bigg[ \left( C_9^{\rm tot} + \frac{\alpha_s C_F}{4 \pi} C_\perp^{(\mbox{\tiny nf,} 9)}(q^2) \right) \nonumber \\
& + \frac{\hat{m}_b}{q^2} \left( (m_B + m_{K^*})\frac{T_1(q_2)}{V(q^2)} + (m_B - m_{K^*}) \frac{T_2(q^2)}{A_1(q^2)} \right)\times \left( C_7^{\mbox{\tiny eff}} + \frac{\alpha_s C_F}{4 \pi}C_\perp^{(\mbox{\tiny nf,} 7)}(q^2) \right) \nonumber \\
& + \frac{\hat{m}_b}{q^2} \left( (m_B + m_{K^*})\frac{1}{V(q^2)} + (m_B - m_{K^*}) \left( 1 - \frac{q^2}{m_B^2} \right) \frac{1}{A_1(q^2)} \right) \nonumber \\
&\times \frac{\alpha_s C_F}{4 \pi} \frac{\pi^2}{N_c} \frac{f_B f_{K^*, \perp} \lambda^{-1}_{B, +}}{m_B} \int_0^1 \dif u \Phi_{K^*, \perp}(u) T_{\perp, +}^{(\mbox{\tiny nf})}(u) \bigg]
\end{align}
where $\Phi_{K^*, \perp}$ is the transverse twist-2 DA for $K^*$. The NLO contribution in Eq. \ref{AFB} is directly sensitive to this DA and therefore it would be interesting to examine its relative significance. 

\begin{figure}
\includegraphics[width = 0.8\textwidth, height = 0.5\textwidth]{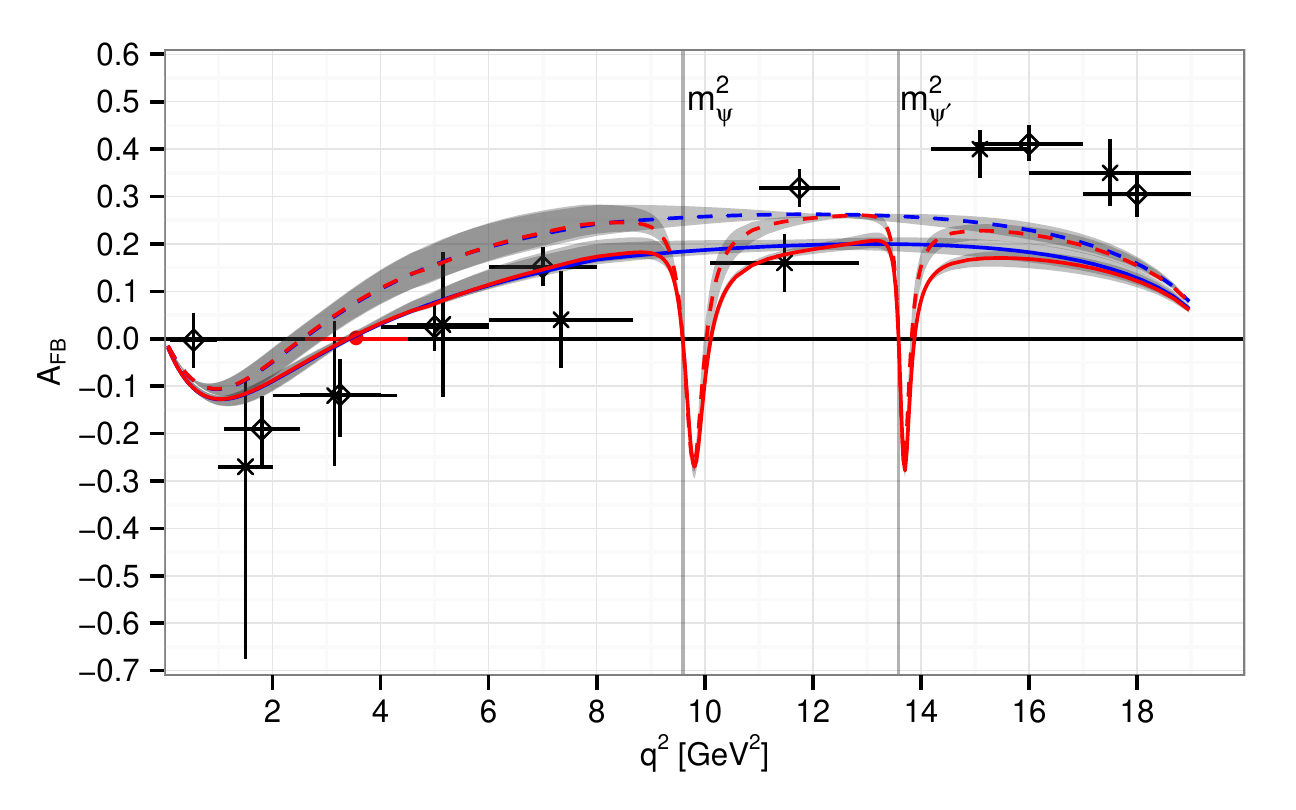}
\caption{LO (dashed) and NLO (solid) predictions for $A_{FB}$ including (red) and excluding (blue) the resonance effects. We compare to the latest LHCb (diamonds) and CMS (crosses) data.  Note that this plot is qualitative and our predictions for each experimental bin for this observable are shown in Tab.\ \ref{tab:bin_AFB} in Appendix A.}
\label{fig:FB_order}
\end{figure}

Our prediction for $A_{\rm FB}$ distribution is given in Fig. \ref{fig:FB_order} in which the latest data points from LHCb, including the zero-crossing point $q_0^2=3.7^{+0.8}_{-1.1}\; {\rm GeV}^2$\cite{LHCb:2015dla}, and CMS \cite{Khachatryan:2015isa} are shown as well.

%%%%%%%%%%%%%%%%%%
\section{Results}%
%%%%%%%%%%%%%%%%%%

\begin{figure}
\vspace{-1.5cm}
\centering
\subfigure[\mbox{ }The differential branching ratio using $q^2$-dependent (solid) and $q^2$-independant (dashed) $f_V$.]{\includegraphics[width=0.40\textwidth]{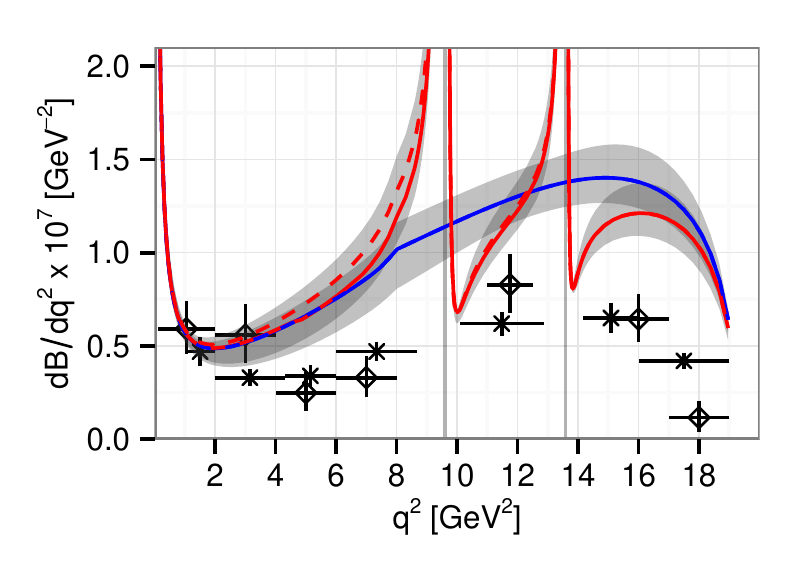}}
\subfigure[\mbox{ }$A_{FB}$ using $q^2$-dependent (solid) and $q^2$-independant (dashed) $f_V$.]{\includegraphics[width=0.40\textwidth]{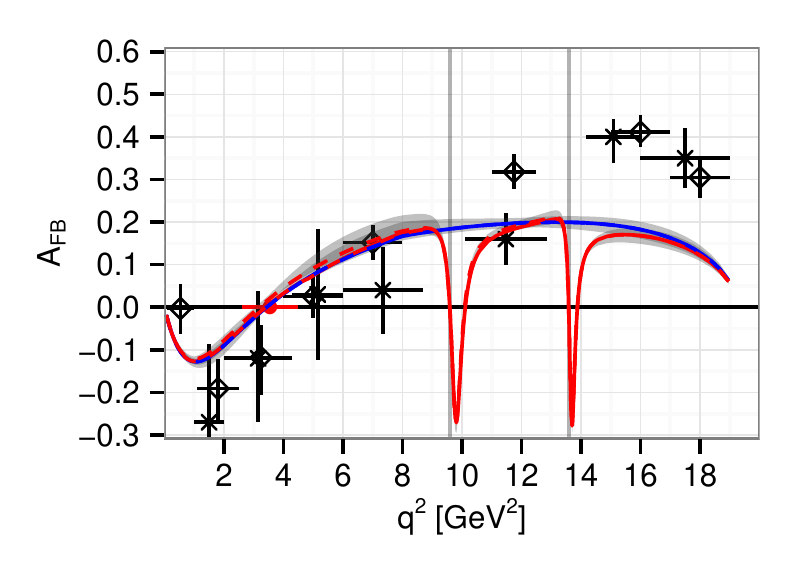}}\\
\subfigure[\mbox{ }The differential branching ratio within the SM (solid) and with new physics $(C_9^{\rm NP}, C_7^{\rm NP}) = (-1.0, -0.01)$ (dashed). The bin by bin predictions are given in the Appendix A.]{\includegraphics[width=0.40\textwidth]{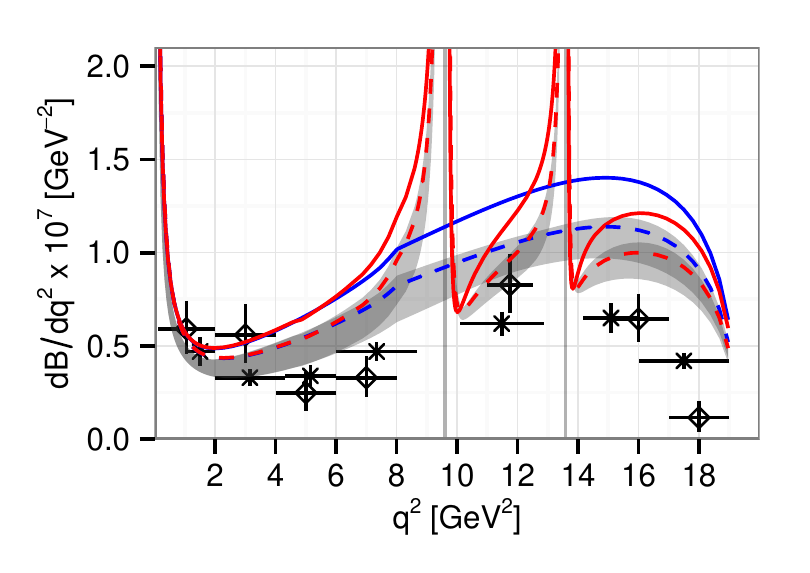}}
\subfigure[\mbox{ } $A_{FB}$ within the SM (solid) and with new physics $(C_9^{\rm NP}, C_7^{\rm NP}) = (-1.0, -0.01)$ (dashed).  The bin by bin predictions are given in the Appendix A.]{\includegraphics[width=0.40\textwidth]{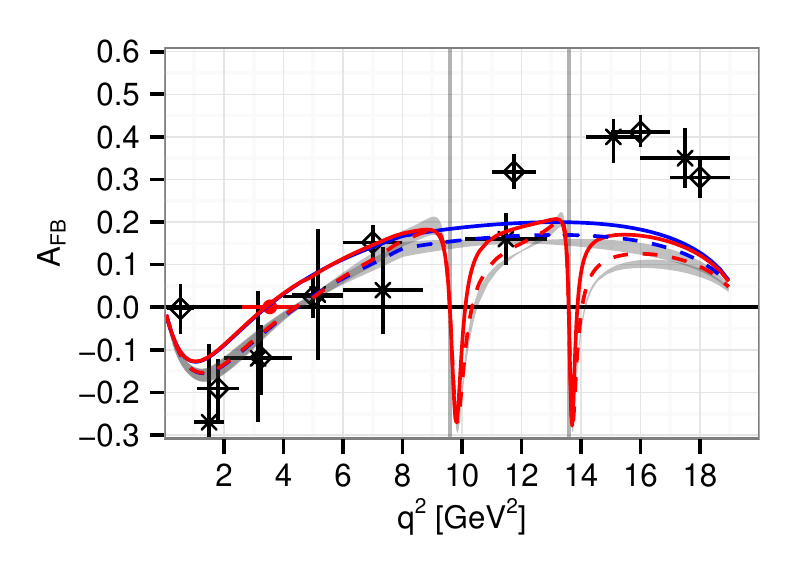}}\\
\subfigure[\mbox{ }The differential branching ratio using AdS/QCD (solid) and SR (dashed) DAs. ]{\includegraphics[width=0.40\textwidth]{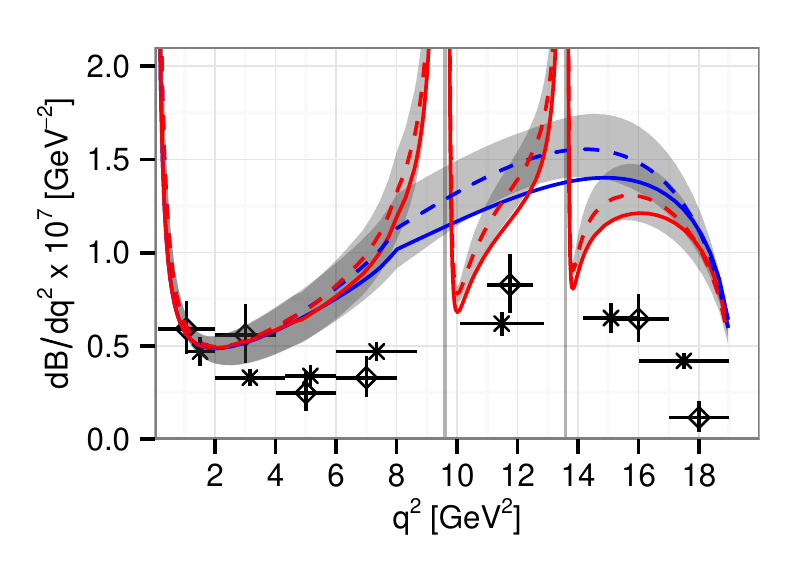}}
\subfigure[\mbox{ }$A_{FB}$ using AdS/QCD (solid) and SR (dashed) DAs.]{\includegraphics[width=0.40\textwidth]{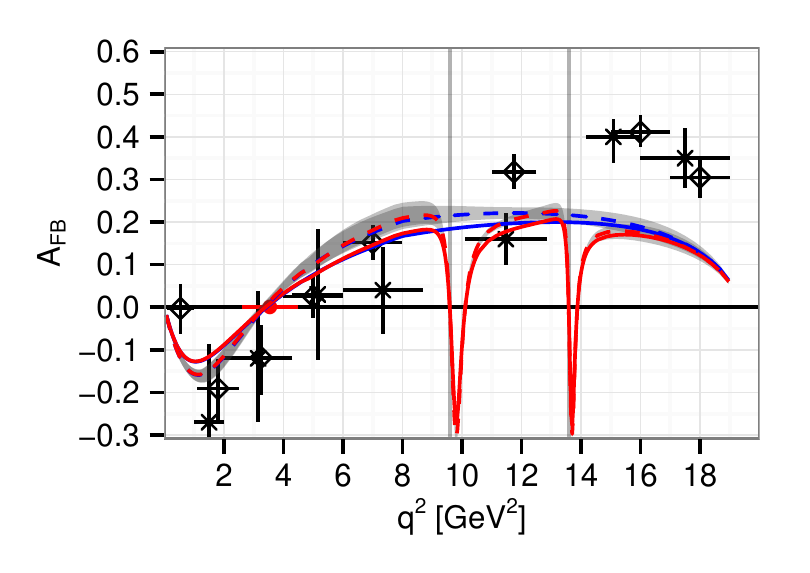}}\\
\caption{Variations in the AdS/QCD predictions of the differential branching ratio and $A_{FB}$ as explained in each figure caption. The red and blue curves show the results with and without the inclusion of $\psi$ and $\psi^\prime$ resonances.}
\label{fig:plot_variations}
\end{figure}

The AdS/QCD predictions for the \process differential branching ratio are shown in Fig. \ref{fig:BR_main}.  We can see that the resonance effects are significant and improve the agreement with the experimental data for $q^2$ regions above $m_\psi^2$.  The gray bands in this figure (and in all subsequent figures) represent the uncertainty due to the renormalization scale $\mu$ (taken in the range $m_b/2\leq\mu \leq 2m_b$) and the error bars on the lattice data for the form factors.  The latter is dominated by the uncertainty in $A_2$ lattice calculations.  Fig.\ \ref{fig:plot_variations}(a) shows our prediction for the differential branching ratio when we assume a momentum-independent decay constant for $\psi$ and $\psi^\prime$ (dashed curve). We note from this graph that the only significant difference occurs at $q^2$ below the first resonance.  As is the case for the inclusive $B\to X_s\ell^+\ell^-$\cite{Ahmady:1995st}, assuming momentum-dependent decay constants leads to better agreement with the experimental data for small $q^2$.  Fig.\ \ref{fig:plot_variations}(c), on the other hand, shows our predictions for the differential branching ratio when additional NP contributions are added to the Wilson coefficients $C_{7}^{\rm eff}$ and $C_{9}^{\rm eff}$.  We note that assuming $C_9^{\rm NP}=-1.0$ and $C_7^{\rm NP} = -0.01$, as suggested by the authors of Ref. \cite{Descotes-Genon:2015xqa}, produces better agreement with the data, especially at high $q^2$. In Fig.\ \ref{fig:plot_variations}(e), we compare our predictions with those obtained from sum rules (SR) DAs.  It seems that AdS/QCD DAs produce results generally lower than those obtained from SR DAs\cite{Ball:2007zt}, especially for larger $q^2$. The predictions for each experimental bin for this observable are shown in Tab.\ \ref{tab:bin_BR} in the Appendix.

Our predictions for $A_{FB}$ are shown in Fig.\ \ref{fig:FB_order}.  First, we observe that 
the leading-order predictions miss all but one of the experimental data points as well as the zero-crossing point.  Second, as pointed out in Ref. \cite{Beneke:2001at}, the inclusion of NLO contributions leads to a significant shift to the zero-crossing point (of order 30\%) and an overall better agreement with the most recent data on $A_{FB}$ below the first resonance.
We observe that the inclusion of the two resonances does not have any noticeable effects for this observable outside the resonance regions. Consequently, as shown in Fig.\ \ref{fig:plot_variations}(b), assuming momentum-independent decay constants for $\psi$ and $\psi^\prime$ does not change our predictions significantly.  On the other hand, assuming the NP contributions $C_9^{\rm NP}=-1.0$ and $C_7^{\rm NP} = -0.01$, produces much better agreement with the experimental data, as seen in Fig.\ \ref{fig:plot_variations}(d). Finally, predictions for $A_{FB}$ based on AdS/QCD DAs are more or less in similar agreement with the data as those obtained from SR DAs, as illustrated in Fig.\ \ref{fig:plot_variations}(f). The predictions for each experimental bin for this observable are shown in Tab.\ \ref{tab:bin_AFB} in the Appendix.

%\begin{figure}
%\includegraphics[width = \textwidth, height = 0.5\textwidth]{plot_FB_NP_chi2_avg_color_3.pdf}
%\caption{The results of our search in the $(C_9^{\rm NP}, C_7^{\rm NP})$ parameter space. The shade intensity indicates the maximal distance of our prediction to the central values of the data points in units of the associated error bars. The green point corresponds to the minimum distance ($0.51$) and the cells identified with back borders correspond to the parameter region where the maximum distance is less than $0.55$.}
%\label{fig:plot_AFB_ps}
%\end{figure}

%%%%%%%%%%%%%%%%%%%%%
\section{Conclusion}%
%%%%%%%%%%%%%%%%%%%%%

We used the form factors and DAs as predicted by the AdS/QCD correspondence as well as taken into account the possible $c\overline{c}$ resonance contributions to give predictions for the \process differential branching ratio and forward-backward asymmetry.  The inclusion of $\psi$ and $\psi^\prime$ resonances is done by using the Breit-Wigner model with momentum-dependent decay constants.  This leads to better agreement with the experiment data for the differential decay rate outside the resonance regions.  However, the forward-backward asymmetry outside the resonance region is not affected by the presence of resonances. We confirm that a negative contribution to $C_9$ and a small contribution to $C_7$, as suggested in Refs \cite{Descotes-Genon:2013wba,Descotes-Genon:2015xqa}, leads to better agreement with the experimental data. Comparison of predictions from AdS/QCD DAs and SR DAs shows that the former produces better or identical results when compared with experimental data on the branching ratio and $A_{FB}$. It would be interesting to investigate the use of our AdS/QCD form factors and DAs to compute other angular observables associated with \process decay for the whole range of $q^2$, in particular, the observable $P_5^\prime$ for which there is a discrepancy between the theory predictions and the LHCb measurement\cite{DescotesGenon:2012zf}.

%%%%%%%%%%%%%%%%%%%%%%%%%%%% 
\section{Acknowledgements} %
%%%%%%%%%%%%%%%%%%%%%%%%%%%%

This research is supported by a team grant from the Natural Sciences and Engineering Research Council of Canada (NSERC).  DH and SL thank the government of New-Brunswick for SEED-COOP funding. The authors would also like to thank the Advance Internet Technologie Research Group (GRETI) and Andy Couturier at l'Universit\'e de Moncton for their technical support. We also thank Alec Morrison for useful discussions.

%%%%%%%%%%
\appendix%
%%%%%%%%%%

\section{Numerical inputs and bin by bin results}

Throughout our analysis, we have used the input parameters presented in Tab.\ \ref{tab:inputs} where all quark, meson and the intermediate boson masses, as well as the experimental value of $\alpha_s(m_Z)$, are taken from the latest Review of Particle Physics \cite{RPP2014}. The two $K^{*}$ decay constants, $f_{K^*}$ and $f_{K^*}^{\perp}$, are AdS/QCD predictions which are dependent on the masses of the quarks in the $K^{*}$ meson \cite{Ahmady:2014sva}.

We use the next-to-next-to leading order evolution for the strong coupling constant $\alpha_s$ which can be found in the Appendix of Ref. \cite{Ahmady:2006yr}.  We also present the values of the $10$ Wilson Coefficients at scale $\mu = m_b$ in Tab.\ \ref{tab:Wilson_coef}. The complete set of equations used to obtain these values have been collected in the appendix of Ref. \cite{Ahmady:2014cpa}.
\begin{table}
	\begin{tabular}{|l||l|}
		\hline
		$m_q = 0.35\mbox{ GeV}$ &	$m_B = 5.28\mbox{ GeV}$ \\
		$m_s = 0.48\mbox{ GeV}$	& 	$m_{K^{*}} = 0.89\mbox{ GeV}$\\
		$m_c = 1.4\mbox{ GeV}$	&	$m_\psi = 3.10\mbox{ GeV}$\\
		$m_b = 4.6\mbox{ GeV}$	&	$m_{\psi^{'}} = 3.69\mbox{ GeV}$\\
		$m_t = 173.5\mbox{ GeV}$&	\\
		\hline
	\end{tabular}
	\hspace{0.5cm}
	\begin{tabular}{|l||l|}
		\hline
		$\alpha_s(m_Z) = 0.1185$	& 	$m_Z = 91.19\mbox{ GeV}$ \\
		$\alpha_{em} = 1/133$		&	\\
		\hline\hline
		$f_{K^*}^{\perp} = 0.119$	&	 	$M_B \mbox{(Borel)} = 8\mbox{ GeV}$\\
		$f_{K^*} = 0.225$ 	&		$s_0 = 36\mbox{ GeV}$ \\
		$f_B = 0.18$ &\\
		\hline
	\end{tabular}
	\caption{Numerical values of the input parameters used in our calculations.}
	\label{tab:inputs}
\end{table}

\begin{table}
	\begin{tabular}{| c | c | c | c | c | c | c | c | c | c | c |}
		\hline
		$C_{1}$ & $C_{2}$ & $C_{3}$ & $C_{4}$ & $C_{5}$ & $C_{6}$ & $C_{7}^{eff}$ & $C_{8}^{eff}$ & $C_{9}$& $C_{10}$\\
		\hline \hline
		$-0.148$ & $1.060$ & $0.012$ & $-0.035$ & $0.010$ & $-0.039$ & $-0.307$ & $-0.169$ & $4.238$ & $-4.641$ \\ \hline
		
	\end{tabular}
	\caption{Values of the Wilson coefficients at $\mu = m_b$.}
	\label{tab:Wilson_coef}
\end{table}
\begin{table}
		\begin{center}
			%\textbf{AdS/QCD predictions for the decay constants of $K^*$}
 \centering
 \begin{tabular}{|c|c|c|c|c|c|c|}
 \hline
 \multicolumn{1}{|c||}{$q^2$ Bin (${\rm GeV}^2$)} &
 \multicolumn{1}{c}{$\left\langle\begin{matrix} dB/dq^2\end{matrix}\right\rangle$} &
 \multicolumn{1}{|c}{$\left\langle\begin{matrix}dB^{res}/dq^2\end{matrix}\right\rangle$} &
 \multicolumn{1}{|c}{$\left\langle\begin{matrix}dB^{NP}/dq^2\end{matrix}\right\rangle$} &
 \multicolumn{1}{|c}{$\left\langle\begin{matrix}dB^{NP,res}/dq^2\end{matrix}\right\rangle$} &
 \multicolumn{1}{|c|}{Experiment} &
 \multicolumn{1}{|c|}{Process}\\
 \hline\hline
%%%%%%%%
% LHCb %
%%%%%%%%
 \multicolumn{1}{|c||}{$[0.10-2.00]$}	& $0.776_{+0.076}^{-0.044}$	% Normal
 & $0.778_{+0.075}^{-0.044}$	% + RES
 & $0.767_{+0.078}^{-0.050}$	% + NP
 & $0.768_{+0.079}^{-0.051}$ % + RES + NP
 & $0.592_{-0.170}^{+0.184}$	% LHCb
 & \parbox[t]{5mm}{\multirow{7}{*}{\rotatebox{-90}{\hspace{-0.5cm}$B^+ \to K^{*+}\mu^+ \mu^-$ (LHCb)}}} \\\cline{1-6}
 \multicolumn{1}{|c||}{$[2.00-4.00]$}	& $0.523_{-0.037}^{+0.044}$	% Normal
 & $0.527_{-0.037}^{+0.045}$	% + RES
 & $0.452_{-0.045}^{+0.047}$ % + NP
 & $0.455_{-0.061}^{+0.046}$ % + RES + NP
 & $0.559_{-0.182}^{+0.197}$	% LHCb
 & \\\cline{1-6}
 \multicolumn{1}{|c||}{$[4.00-6.00]$}   & $0.664_{-0.063}^{+0.070}$ % Normal
 & $0.665_{-0.063}^{+0.070}$ % + RES
 & $0.552_{-0.068}^{+0.072}$	% + NP
 & $0.553_{-0.080}^{+0.072}$ % + RES + NP
 & $0.249_{-0.113}^{+0.127}$	% LHCb
 & \\\cline{1-6}
 \multicolumn{1}{|c||}{$[6.00-8.00]$}   & $0.869_{-0.097}^{+0.103}$	% Normal
 & $0.930_{-0.105}^{+0.111}$	% + RES
 & $0.709_{-0.096}^{+0.102}$	% + NP
 & $0.755_{-0.092}^{+0.112}$	% + RES + NP
 & $0.330_{-0.123}^{+0.136}$	% LHCb
 & \\\cline{1-6}
 \multicolumn{1}{|c||}{$[11.00-12.50]$} & $1.286_{-0.133}^{+0.141}$ % Normal
 & $1.174_{-0.121}^{+0.128}$	% + RES
 & $1.043_{-0.130}^{+0.137}$	% + NP
 & $0.966_{-0.081}^{+0.117}$	% + RES + NP
 & $0.828_{-0.197}^{+0.214}$	% LHCb
 & \\\cline{1-6}
 \multicolumn{1}{|c||}{$[15.00-17.00]$} & $1.370_{-0.123}^{+0.135}$	% Normal
 & $1.198_{-0.105}^{+0.116}$	% + RES
 & $1.114_{-0.138}^{+0.145}$	% + NP
 & $0.990_{-0.081}^{+0.118}$	% + RES + NP
 & $0.644_{-0.159}^{+0.173}$	% LHCb
 & \\\cline{1-6}
 \multicolumn{1}{|c||}{$[17.00-19.00]$} & $1.072_{-0.125}^{+0.030}$	% Normal
 & $0.984_{-0.114}^{+0.029}$	% + RES
 & $0.872_{-0.050}^{+0.027}$	% + NP
 & $0.807_{-0.051}^{+0.026}$	% + RES + NP
 & $0.116_{-0.084}^{+0.099}$	% LHCb
 & \\\cline{1-6}
 \hline\hline
 
 %%%%%%%
 % CMS %
 %%%%%%%
 \multicolumn{1}{|c||}{$[1.00-2.00]$}	& $0.510_{+0.010}^{-0.005}$	% Normal
 & $0.512_{+0.010}^{-0.005}$	% + RES
 & $0.473_{+0.024}^{-0.009}$	% + NP
 & $0.475_{+0.023}^{-0.008}$ % + RES + NP
 & $0.47_{-0.076}^{+0.076}$	% CMS
 & \parbox[t]{5mm}{\multirow{7}{*}{\rotatebox{-90}{\hspace{-0.5cm}$B^0 \to K^{*0}\mu^+ \mu^-$ (CMS)}}} \\\cline{1-6}
 \multicolumn{1}{|c||}{$[2.00-4.30]$}	& $0.532_{-0.040}^{+0.046}$	% Normal
 & $0.535_{-0.040}^{+0.047}$	% + RES
 & $0.458_{-0.030}^{+0.020}$ % + NP
 & $0.461_{-0.030}^{+0.020}$ % + RES + NP
 & $0.33_{-0.045}^{+0.045}$	% CMS
 & \\\cline{1-6}
 \multicolumn{1}{|c||}{$[4.30-6.00]$}   & $0.676_{-0.053}^{+0.065}$ % Normal
 & $0.677_{-0.053}^{+0.065}$ % + RES
 & $0.561_{-0.045}^{+0.052}$	% + NP
 & $0.562_{-0.045}^{+0.052}$ % + RES + NP
 & $0.34_{-0.058}^{+0.058}$	% CMS
 & \\\cline{1-6}
 \multicolumn{1}{|c||}{$[6.00-8.68]$}   & $0.914_{-0.100}^{+0.106}$	% Normal
 & $1.031_{-0.115}^{+0.120}$	% + RES
 & $0.743_{-0.082}^{+0.089}$	% + NP
 & $0.834_{-0.082}^{+0.095}$	% + RES + NP
 & $0.47_{-0.050}^{+0.050}$	% CMS
 & \\\cline{1-6}
 \multicolumn{1}{|c||}{$[10.09-12.86]$} & $1.396_{-0.121}^{+0.138}$ % Normal
 & $1.148_{-0.094}^{+0.109}$	% + RES
 & $1.027_{-0.091}^{+0.106}$	% + NP
 & $0.922_{-0.063}^{+0.078}$	% + RES + NP
 & $0.62_{-0.064}^{+0.064}$	% CMS
 & \\\cline{1-6}
 \multicolumn{1}{|c||}{$[14.18-16.00]$} & $1.164_{-0.115}^{+0.300}$ 	% Normal
 & $1.058_{-0.104}^{+0.290}$	% + RES
 & $1.135_{-0.089}^{+0.027}$	% + NP
 & $0.960_{-0.079}^{+0.027}$	% + RES + NP
 & $0.65_{-0.078}^{+0.078}$	% CMS
 & \\\cline{1-6}
 \multicolumn{1}{|c||}{$[16.00-19.00]$} & $1.164_{-0.115}^{+0.300}$ 	% Normal
 & $1.058_{-0.104}^{+0.290}$	% + RES
 & $0.946_{-0.089}^{+0.027}$	% + NP
 & $0.868_{-0.079}^{+0.027}$	% + RES + NP
 & $0.42_{-0.042}^{+0.042}$	% CMS
 & \\\cline{1-6}
 \hline

 \end{tabular}
		\end{center}
		\caption {Bin by bin values of the branching ratio defined as $\frac{10^{7}}{q_{2}^2-q_{1}^2}\times\int_{q_1^2}^{q_2^2}\frac{dB}{dq^{2}}dq^2$ for $[q_1^2,q_2^2]$ bin, with and without resonances as well as new physics contributions as compared with the latest LHCb\cite{Aaij:2014pli} and CMS \cite{Khachatryan:2015isa} data}.
		\label{tab:bin_BR}
	\end{table}
\begin{table}
		\begin{center}
 \centering
 \begin{tabular}{|c|c|c|c|c|c|c|}
 \hline
 \multicolumn{1}{|c||}{$q^2$ Bin (${\rm GeV}^2$)} &
 \multicolumn{1}{c}{$\left\langle\begin{matrix} A_{FB}\end{matrix}\right\rangle$} &
 \multicolumn{1}{|c}{$\left\langle\begin{matrix} A_{FB}^{res}\end{matrix}\right\rangle$} &
 \multicolumn{1}{|c}{$\left\langle\begin{matrix} A_{FB}^{NP}\end{matrix}\right\rangle$} &
 \multicolumn{1}{|c}{$\left\langle\begin{matrix} A_{FB}^{res,NP}\end{matrix}\right\rangle$} &
 \multicolumn{1}{|c}{Experiment} &
 \multicolumn{1}{|c|}{Process} \\
 \hline\hline
%%%%%%%%
% LHCb %										
%%%%%%%%								
 \multicolumn{1}{|c||}{$[0.10-0.98]$}	& $-0.096_{+0.010}^{-0.003}$	% Normal
 & $-0.095_{+0.010}^{-0.003}$	% + RES
 & $-0.103_{+0.012}^{-0.005}$	% + NP
 & $-0.103_{+0.010}^{-0.005}$ 	% + RES + NP
 & $-0.003_{-0.060}^{+0.058}$	% LHCb
 & \parbox[t]{5mm}{\multirow{8}{*}{\rotatebox{-90}{\hspace{-0.6cm}$B^0 \to K^{*0}\mu^+ \mu^-$ (LHCb)}}} \\\cline{1-6}
 \multicolumn{1}{|c||}{$[1.10-2.50]$}	& $-0.099_{-0.016}^{+0.011}$	% Normal
 & $-0.097_{-0.017}^{+0.011}$	% + RES
 & $-0.140_{-0.020}^{+0.010}$	% + NP
 & $-0.138_{-0.021}^{+0.010}$	% + RES + NP
 & $-0.191_{-0.079}^{+0.070}$	% LHCb
 & \\\cline{1-6}
 \multicolumn{1}{|c||}{$[2.50-4.00]$}	& $-0.011_{-0.005}^{+0.010}$	% Normal
 & $-0.010_{-0.005}^{+0.010}$	% + RES
 & $-0.065_{-0.017}^{+0.014}$	% + NP
 & $-0.063_{-0.017}^{+0.014}$	% + RES + NP
 & $-0.118_{-0.088}^{+0.075}$	% LHCb
 & \\\cline{1-6}
 \multicolumn{1}{|c||}{$[4.00-6.00]$} 	& $0.075_{-0.020}^{+0.005}$		% Normal
 & $0.075_{-0.020}^{+0.005}$		% + RES
 & $0.025_{-0.009}^{+0.010}$		% + NP
 & $0.025_{-0.007}^{+0.010}$		% + RES + NP
 & $0.025_{-0.050}^{+0.050}$		% LHCb
 & \\\cline{1-6}
 \multicolumn{1}{|c||}{$[6.00-8.00]$} 	& $0.141_{+0.029}^{-0.007}$ 	% Normal
 & $0.146_{+0.031}^{-0.012}$		% + RES
 & $0.103_{-0.018}^{+0.009}$		% + NP
 & $0.113_{-0.019}^{+0.006}$		% + RES + NP
 & $0.152_{-0.041}^{+0.041}$		% LHCb
 & \\\cline{1-6}
 \multicolumn{1}{|c||}{$[11.00-12.50]$}	& $0.197_{+0.012}^{-0.011}$ 	% Normal
 & $0.182_{+0.020}^{-0.015}$		% + RES
 & $0.153_{-0.005}^{+0.005}$		% + NP
 & $0.140_{-0.005}^{+0.004}$		% + RES + NP
 & $0.318_{-0.041}^{+0.041}$		% LHCb
 & \\\cline{1-6}
 \multicolumn{1}{|c||}{$[15.00-17.00]$}	& $0.181_{+0.019}^{-0.015}$ 	% Normal
 & $0.166_{+0.017}^{-0.014}$		% + RES
 & $0.104_{+0.012}^{-0.012}$		% + NP
 & $0.123_{+0.009}^{-0.008}$		% + RES + NP
 & $0.411_{-0.036}^{+0.041}$		% LHCb
 & \\\cline{1-6}
 \multicolumn{1}{|c||}{$[17.00-19.00]$}	& $0.123_{+0.012}^{-0.012}$ 	% Normal
 & $0.117_{+0.011}^{-0.011}$		% + RES
 & $0.082_{+0.009}^{-0.010}$		% + NP
 & $0.091_{+0.008}^{-0.009}$		% + RES + NP
 & $0.305_{-0.050}^{+0.049}$		% LHCb
 & \\\cline{1-6}
 \hline\hline
 %%%%%%%
 % CMS %
 %%%%%%%
 \multicolumn{1}{|c||}{$[1.00-2.00]$}	& $-0.114_{+0.016}^{-0.010}$	% Normal
 & $-0.113_{+0.016}^{-0.010}$	% + RES
 & $-0.150_{+0.010}^{-0.007}$	% + NP
 & $-0.148_{+0.011}^{-0.009}$ 	% + RES + NP
 & $-0.27_{-0.406}^{+0.184}$		% CMS
 & \parbox[t]{5mm}{\multirow{7}{*}{\rotatebox{-90}{\hspace{-0.5cm}$B^0 \to K^{*0}\mu^+ \mu^-$ (CMS)}}} \\\cline{1-6}
 \multicolumn{1}{|c||}{$[2.00-4.30]$}	& $-0.018_{-0.006}^{+0.010}$	% Normal
 & $-0.017_{-0.006}^{+0.010}$	% + RES
 & $-0.071_{-0.017}^{+0.013}$	% + NP
 & $-0.069_{-0.016}^{+0.013}$	% + RES + NP
 & $-0.12_{-0.149}^{+0.158}$		% CMS
 & \\\cline{1-6}
 \multicolumn{1}{|c||}{$[4.3-6.00]$}	& $0.081_{-0.022}^{+0.005}$	% Normal
 & $0.081_{-0.022}^{+0.005}$	% + RES
 & $0.032_{-0.011}^{+0.010}$	% + NP
 & $0.032_{-0.011}^{+0.010}$	% + RES + NP
 & $0.03_{-0.153}^{+0.153}$	% CMS
 & \\\cline{1-6}
 \multicolumn{1}{|c||}{$[6.00-8.68]$} 	& $0.149_{-0.029}^{+0.007}$		% Normal
 & $0.154_{-0.030}^{+0.008}$		% + RES
 & $0.113_{-0.022}^{+0.006}$		% + NP
 & $0.126_{-0.027}^{+0.006}$		% + RES + NP
 & $0.04_{-0.101}^{+0.101}$		% CMS
 & \\\cline{1-6}
 \multicolumn{1}{|c||}{$[10.09-12.86]$} & $0.195_{+0.012}^{-0.011}$ 	% Normal
 & $0.161_{+0.005}^{-0.005}$		% + RES
 & $0.165_{-0.014}^{+0.011}$		% + NP
 & $0.109_{-0.019}^{+0.006}$		% + RES + NP
 & $0.16_{-0.061}^{+0.061}$		% CMS
 & \\\cline{1-6}
 \multicolumn{1}{|c||}{$[14.18-16.00]$}	& $0.192_{+0.019}^{-0.015}$ 	% Normal
 & $0.163_{+0.014}^{-0.011}$		% + RES
 & $0.163_{-0.010}^{+0.008}$		% + NP
 & $0.109_{-0.007}^{+0.008}$		% + RES + NP
 & $0.40_{-0.061}^{+0.041}$		% CMS
 & \\\cline{1-6}
 \multicolumn{1}{|c||}{$[16.00-19.00]$}	& $0.140_{+0.014}^{-0.013}$ 	% Normal
 & $0.131_{+0.013}^{-0.012}$		% + RES
 & $0.118_{+0.010}^{-0.010}$		% + NP
 & $0.102_{+0.007}^{-0.021}$		% + RES + NP
 & $0.35_{-0.071}^{+0.071}$		% CMS
 & \\\cline{1-6}
 \hline

 \end{tabular}
		\end{center}
		\caption {Bin by bin values of $A_{FB}$ defined as$\frac{1}{q_{2}^2-q_{1}^2}\times\int_{q_1^2}^{q_2^2}\frac{d A_{FB}}{dq^{2}}dq^2$, with and without resonances as well as new physics contributions as compared with the latest LHCb  \cite{LHCb:2015dla} and CMS \cite{Khachatryan:2015isa} data.}
		\label{tab:bin_AFB}
	\end{table}

%%%%%%%%%%%%%%%%
% Bibliography %
%%%%%%%%%%%%%%%%

\bibliographystyle{apsrev}
\bibliography{kstarFB_PRD}

\end{document}